\documentclass[reprint,aps,pre]{revtex4-2}

\pdfoutput=1

\usepackage{graphicx}% Include figure files
\usepackage{dcolumn}% Align table columns on decimal point
\usepackage{bm}% bold math

\makeatletter

\usepackage{amsmath}
\usepackage[shortlabels]{enumitem}
\usepackage{etoolbox}
\usepackage[utf8]{inputenc}
\usepackage{mathtools}
\usepackage[activate={true,nocompatibility},
	final,
	tracking=true,
	kerning=true,
	spacing=true,
	factor=1100,
	stretch=10,
	shrink=10]{microtype}
 
\usepackage{pgfplots}
\usepackage[caption=false]{subfig}

\usepackage{tabularx}
\usepackage{booktabs}

\microtypecontext{spacing=nonfrench}  % silence warning

\AtBeginDocument{
 
 \newcommand{\dfr}[3][]{\frac{d^{#1} #2}{d #3{}^{#1}}}
 \newcommand{\@dpfr}[3][]{\frac{\partial^{#1} #2}{\partial #3{}^{#1}}}
 \newcommand{\@spfr}[3][]{\partial^{#1} #2 / \partial #3{}^{#1}}
 \newcommand{\pfr}[3][]{\mathchoice{\@dpfr[#1]{#2}{#3}}{\@dpfr[#1]{#2}{#3}}
  {\@spfr[#1]{#2}{#3}}{\@spfr[#1]{#2}{#3}}}
}
\DeclarePairedDelimiter\paren{(}{)}

\DeclarePairedDelimiter\abs{\lvert}{\rvert}

\DeclarePairedDelimiter\bkt{[}{]}
\DeclarePairedDelimiter\set{\{}{\}}
\let\oldparen\paren
\def\paren{\@ifstar{\oldparen}{\oldparen*}}
\let\oldbkt\bkt
\def\bkt{\@ifstar{\oldbkt}{\oldbkt*}}

\usepackage[hidelinks]{hyperref}% add hypertext capabilities
\usepackage[capitalize]{cleveref}
\appto{\appendix}{%
  \@ifstar{\def\theequation@prefix{A.}}%
          {}%
}
\makeatother

\begin{document}

\preprint{APS/123-QED}

\title{Interacting systems with zero thermodynamic curvature}

\author{Juan Rodrigo}
\email{jrodrigo@nip.upd.edu.ph}
\author{Ian Vega}%
\affiliation{%
	National Institute of Physics, University of the Philippines, Diliman,
	Quezon City, 1101, Philippines
}%

\date{\today}

\begin{abstract}

We review a conjecture by Ruppeiner that relates the nature of interparticle
interactions to the sign of the thermodynamic curvature scalar $R$, paying special
attention to the case of zero curvature. We highlight the underappreciated 
fact that there are two Ruppeiner metrics that are equally viable in principle,
which are obtained by restricting to systems of constant volume and constant
particle number, respectively. We then demonstrate the existence of
thermodynamic systems with vanishing curvature scalar but nontrivial interactions.
Information about interactions in these systems is obtained
by carrying out an inversion procedure on the virial coefficients.
Finally, we show using the virial expansion that the ideal gas
is the unique physical system for which both curvature scalars vanish.
This leads us to propose an extension to Ruppeiner's conjecture.

% \begin{description}
% \item[Usage]
% Secondary publications and information retrieval purposes.
% \item[Structure]
% You may use the \texttt{description} environment to structure your abstract;
% use the optional argument of the \verb+\item+ command to give the category of each item. 
% \end{description}
\end{abstract}

%Use showkeys class option if keyword
                              %display desired
\maketitle

\section{\label{sec:level1}Introduction}

There is a long history of geometric ideas applied to thermodynamics.
An early advocate was J.~W.~Gibbs, who envisioned the equilibrium states of a system
as existing on a ``thermodynamic surface''~\cite{gibbs1928collected}. This viewpoint
trades an equation of state and its derivatives 
for the thermodynamic surface and its associated tangents~\cite{callen1985thermodynamics}.
Much later, it would be shown that this surface (and closely related versions of it)
could be used to define a metric on a system's state space,
thereby imbuing thermodynamics with the geometric concepts of length, geodesics,
and curvature~\cite{weinhold1975metric,weinhold1975metricii,ruppeiner1979thermodynamics}.
These ideas have since gone beyond the initial scope of
equilibrium thermodynamics to find use in the study of finite-time thermodynamics
and microscopic interactions, to name just a few applications.

The metric we study in this work was proposed by Ruppeiner~\cite{ruppeiner1979thermodynamics},
who showed that thermodynamic fluctuation theory comes with a natural
metric structure, in the sense that the distances defined by this metric are
directly interpretable in terms of the probabilities of fluctuations. A significant
body of work has grown out of the study of the Ruppeiner
metric~\cite{ruppeiner1995riemannian}; we provide a brief survey here.

Arguably the foundational result in Ruppeiner geometry is the fact that
the ideal gas possesses a flat metric. This result led to the conjecture that the
curvature scalar $R$ derived from the metric is
proportional to interactions:
\begin{equation}
	\abs{R} \sim \xi^{d},
	\label{eq:Rsimxi}
\end{equation}
where $\xi$ is the correlation length and $d$ is the number of spatial dimensions
of the system. This proportionality was also confirmed
in the asymptotic critical region for a number of systems
\cite{ruppeiner1979thermodynamics,ruppeiner1981application,ruppeiner1990thermodynamic},
and later work, based on the formalism of
path integrals, would provide more rigorous underpinning for this
conjecture~\cite{ruppeiner1983new,ruppeiner1983thermodynamic,diosi1985covariant}.
Thus, the metric seems to provide
more than just a notion of distance, and appears to reveal information about
the micro- and mesoscopic regimes of a system. Much work in Ruppeiner geometry involves
exploring this relationship between interactions and
curvature~\cite{ruppeiner2010thermodynamic}.

Thermodynamic geometry has also been used to study
phase transitions in thermodynamic
systems~\cite{diosi1989mapping,de2022phase,ruppeiner2012thermodynamic,ruppeiner2012critical,may2012riemannian}. It has been shown that the envelope obtained
by propagating geodesics of the Ruppeiner metric in state space can be used to
partition the phases of the van der Waals
fluid~\cite{diosi1989mapping,de2022phase}. Further, these geodesics appear to
detect the presence of the Widom line, which is believed by some to be the
extension of the phase coexistence curve into the supercritical
region~\cite{gorelli2006liquidlike,simeoni2010widom}. Ruppeiner geometry was also
used to construct the Widom line for realistic fluids using
multiparameter equations of state constructed from experimental data and computer
simulations of the Lennard-Jones fluid~\cite{ruppeiner2012thermodynamic,may2012riemannian}.
Another construction based on the curvature scalar, the so-called $R$-crossing
method~\cite{ruppeiner2012critical,ruppeiner2012thermodynamic,de2022phase}, has been
put forward as an alternative to the Maxwell construction for the phase coexistence curve.

A particularly intriguing conjecture in Ruppeiner geometry emerged from
the study of the curvature scalar $R$ for model systems where the interactions are
known~\cite{ruppeiner2010thermodynamic,janyszek1990riemannian,may2013thermodynamic,branka2018thermodynamic,ruppeiner2021thermodynamic,mirza2008ruppeiner,mirza2009nonperturbative}.
This conjecture states that that the sign of $R$ indicates whether interactions
are primarily attractive or repulsive. (Interpretation of the sign depends on which
convention is used for the calculation of the curvature scalar.)
It has been shown that $R<0$ for the
ideal Bose gas and $R > 0$ for the ideal Fermi gas, where quantum statistics mimic
the effects of attractive and repulsive interactions,
respectively~\cite{janyszek1990riemannian}. This conjecture
has also been probed more directly using Weeks-Chandler-Andersen perturbation
theory~\cite{may2013thermodynamic}. More recently, it has been discovered that hard-sphere
gases tend to possess a curvature scalar which is positive in some regions and
negative in others, seemingly contradicting the fact that hard-sphere gases interact
strictly via repulsion~\cite{branka2018thermodynamic,ruppeiner2021thermodynamic}.

Thermodynamic geometry has also been extended to the study of black hole
thermodynamics~\cite{ruppeiner2008thermodynamic,aaman2007ruppeiner,mirza2007ruppeiner,wei2019repulsive,wei2019ruppeiner}.
Interpretation of the Ruppeiner metric here requires a bit more care,
due to the existence of nontrivial black hole solutions with flat thermodynamic
metrics~\cite{medved2008commentary}. Other authors have carried over Ruppeiner's
conjecture in order to relate the thermodynamic curvature scalar and black
hole microstructure~\cite{wei2019repulsive,wei2019ruppeiner}.

The concept of thermodynamic length
has also proven to be valuable in finite-time thermodynamics, where
it provides bounds on dissipation and entropy
production~\cite{salamon1983thermodynamic,salamon1985length}.
In particular, geodesics in thermodynamic geometry have been shown to
correspond to paths of minimum entropy production or
dissipation~\cite{nulton1985quasistatic}, which is relevant in the study of optimal
finite-time processes~\cite{bravetti2016thermodynamic,sivak2012thermodynamic}.

In this work, we explore systems with flat thermodynamic metrics, which, according to Ruppeiner geometry, are typically interpreted as indicating the absence of interactions. The Ruppeiner conjecture posits that $R=0$ should correspond uniquely to the ideal gas. However, as we will demonstrate, this is not always the case. This then raises the intriguing question: what kinds of interactions characterize these zero-curvature, non-ideal gases?
While a few papers
have touched on this topic in passing~\cite{aaman2006flat,garcia2014geometric,aaman2015thermodynamic}, 
there has been no comprehensive or systematic investigation of these systems, nor a detailed analysis of their physical significance and implications for Ruppeiner's conjecture. This paper seeks to fill that gap, as we believe that a deeper understanding of these systems is essential, particularly given the increasing use of thermodynamic curvature as a tool for probing interactions.

A key point that we emphasize, often overlooked in the literature, is that different choices of thermodynamic surfaces lead to different Ruppeiner geometries, which, in turn, give distinct classes of interacting systems with zero thermodynamic curvature. In particular, we find approximately flat metrics corresponding to certain inverse-power and hard-sphere potentials. These examples lead us to propose an extension to Ruppeiner's conjecture which appears to single out the ideal gas as the sole example of a non-interacting system.

This paper is organized as follows. Section~\ref{sec:rupp-geom} is a
review of the derivation of the Ruppeiner metric, with particular emphasis on
the choice of induced metric. Section~\ref{sec:vanishing-curvature} concerns
systems with vanishing thermodynamic curvature, i.e., systems for which
the thermodynamic curvature scalar satisfies $R = 0$. First
we obtain exact solutions, and then
we propose a different solution-finding method based on the virial expansion of
gases. We then provide interpretation
of the virial-coefficient solutions using an inversion procedure. Finally, we present
our conclusions and recommendations for future work in
Section~\ref{sec:conclusions}.

\section{Ruppeiner geometry}
\label{sec:rupp-geom}

We may define a metric on thermodynamic state space as follows~\cite{ruppeiner1995riemannian}:
\begin{equation}
	g = g_{\mu\nu}\,dX^\mu \,dX^\nu,
	\label{eq:grupp-defn}
\end{equation}
where $X^\mu= (U,V,N)$,
are the natural variables of the entropy $S$, i.e., the internal
energy, volume, and number of particles in the system, respectively.
The metric components $g_{\mu\nu}$ are given by the negative
Hessian of $S$ with respect to these variables,
\begin{equation}
	g_{\mu\nu} = -\frac1k\frac{\partial^2 S}{\partial X^\mu\partial X^\nu},
	\label{eq:grupp-Shessian}
\end{equation}
where $k$ is Boltzmann's constant.
Positive-definiteness is ensured by the principle that entropy is maximized for
equilibrium states, and symmetry follows from the equality of mixed partial
derivatives.
We note that the full metric defined by Eq.~\eqref{eq:grupp-defn} is generally
degenerate due to the first-order homogeneity of the entropy, $S(\lambda X^\mu)
=\lambda S(X^\mu)$. To demonstrate this, we write the Gibbs-Duhem relation
implied by homogeneity (summing over repeated indices):
\begin{equation}
0 = X^i \,dY_i,
\label{eq:gibbsduhem}
\end{equation}
where $Y_i := \partial S / \partial X^i$. Expanding the $dY_i$, we obtain
\begin{equation}
	0 = (D^2S)_{ij}\,X^i \,dX^j,
\end{equation}
and since this relation is true for arbitrary variations $dX^j$, it follows
that the rows of the matrix  $D^2S$ are linearly dependent, which in turn
renders the metric $g$ degenerate. This degeneracy can
be avoided by choosing one of the extensive variables to be fixed, typically
either volume $V$ or particle number $N$. This argument complements the requirement
in the fluctuation-theory derivation of the metric that one extensive
variable must be held fixed in order to set a scale for the
system~\cite{de2022phase,ruppeiner1995riemannian}.
Note that the same is not required to hold in black hole thermodynamics,
where, e.g., all three
variables $(M,L,Q)$, which denote the black hole mass, angular momentum, and charge, respectively,
of a Kerr-Newman black hole can be allowed to
vary~\cite{ruppeiner2008thermodynamic}. The argument using the Gibbs-Duhem relation~\eqref{eq:gibbsduhem} is not applicable here, since black hole entropy $S$ is not
generally extensive in the variables $(M,L,Q)$.

Next, we make a change of variables to eliminate the internal energy $U$ in favor
of temperature $T$, which gives
\begin{equation}
	g = \frac1T\paren{\pfr ST}\,dT^2 + \frac1T\paren{\pfr{P_i}{X^j}}
	\,dX^i\,dX^j,
 \label{eq:g-TXcoords}
\end{equation}
where $P_i := \partial U/\partial X^i$, so that the metric is diagonal in this
coordinate system (see Ref.~\cite{ruppeiner2012critical} for alternative choices).
This equation is more naturally expressed
in terms of the Helmholtz free energy $F$, and with the choice of fixed $V$ we obtain
the metric
\begin{equation}
	g_V = -\frac1T \paren{\pfr[2] FT}\,dT^2 + \frac1T \paren{\pfr[2] FN}
	\, dN^2.
	\label{eq:gV-defn}
\end{equation}
This metric is the most frequently studied in the literature, and is more
commonly written with $N$ replaced by the number density $\rho := N /V$.
This is easily seen to be
equivalent to the expression above.
On the other hand, when $N$ is held constant, we obtain
\begin{equation}
	g_N = -\frac1T \paren{\pfr[2] FT}\,dT^2 + \frac1T \paren{\pfr[2] FV}
	\, dV^2.
	\label{eq:gN-defn}
\end{equation}
We will refer to these as the \emph{Ruppeiner-$V$} and \emph{Ruppeiner-$N$} metrics,
respectively~\cite{de2022phase}. Both Eqs.~\eqref{eq:gV-defn}
and~\eqref{eq:gN-defn} take the form
\begin{equation}
	g_Y = g_{TT}\, dT^2 + g_{XX}\, dX^2,
	\label{eq:ruppmetric-Frep}
\end{equation}
where $X$ can denote either volume $V$ or particle number~$N$, and
$Y$ is the extensive variable being held fixed.
The curvature scalar is then given by
\begin{align}
	R_{Y} &= \frac1{\sqrt{g_{TT}g_{XX}}}\Bigg(\pfr{}T\paren{
		\frac1{\sqrt{g_{TT}g_{XX}}}	\pfr{g_{XX}}T}\nonumber \\[5pt]
&\hspace{70pt}{}+ \pfr{}X\paren{
	\frac1{\sqrt{g_{TT}g_{XX}}}
	\pfr{g_{TT}}X
}\Bigg).
\label{eq:thermocurvature-RY}
\end{align}
In more geometric language, these metrics are induced from the full Ruppeiner
metric on hypersurfaces of constant volume and particle number, respectively.
As was reviewed in Ref.~\cite{de2022phase}, the Ruppeiner-$N$ metric is a
relatively unpopular choice in the literature, but seems more appropriate in,
e.g.,~phase transition analysis, where the number of particles in the system
is held constant. Since both metrics seem equally viable in principle, we shall study
both, being careful to indicate our choice of induced metric with the appropriate
subscript.

For later reference, we also note the possibility of expressing the Ruppeiner metric
in terms of response functions. It is straightforward to see from Eq.~\eqref{eq:g-TXcoords} that
\begin{align}
	g_V &= \frac{C_V}{T^2}\, dT^2 + \frac1T \paren{\pfr\mu N}_{T,V}dN^2,
	\label{eq:gV-respfxns}\\[5pt]
	g_N &= \frac{C_V}{T^2}\, dT^2 + \frac{B_T}{VT}\,dV^2,\label{eq:gN-respfxns}
\end{align}
where $C_V := T(\partial S/\partial T)_V$ is the isochoric heat capacity,
$B_T := -V(\partial P/\partial V)_T$ is the isothermal bulk modulus,
 and $\mu$ is the chemical potential. These equations show that knowledge of
 the response functions of a system is sufficient to specify its Ruppeiner geometry.

\subsection{Example: van der Waals}

\begin{figure}[tb]
\centering
\includegraphics[width=0.95\columnwidth]{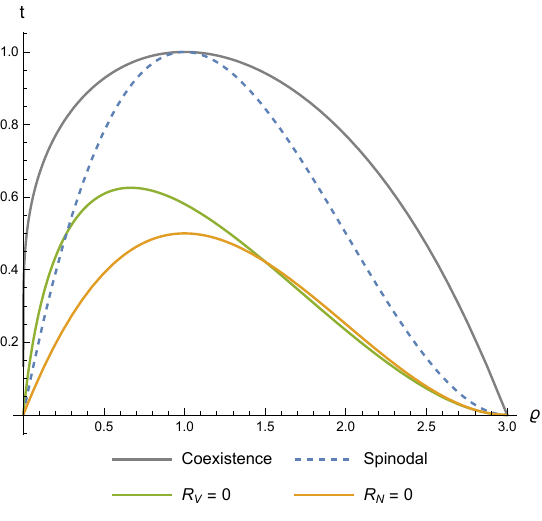}
\caption{Sign-changing curves for the curvature scalars of the van der Waals fluid.
The entire $R_N = 0$ curve lies beneath the spinodal, i.e., within the thermodynamically
unstable region, and is therefore inaccessible to the system.
On the other hand, a portion of the $R_V = 0$ curve near $\varrho = 0$ 
lies above the spinodal and may be accessible by metastable states.}
\label{fig:vdw-rcurves}
\end{figure}

The two metrics~\eqref{eq:gV-defn} and \eqref{eq:gN-defn} generally describe
distinct geometries, as will become clear later.
For an elementary example, we consider the van der Waals gas,
which has a Helmholtz free energy given by
\begin{equation}
F
= - NkT\ln\bkt{\paren{\frac{V}{N} - b} (T\eta)^{3/2}} - \frac{aN^2}{V},
\label{eq:Fvdw}
\end{equation}
where $a$, $b$, and $\eta$ are constants.
We note that a natural temperature and number density scale is
provided by the critical point of the system, which is located at
\begin{equation}
T_c = \frac{8}{27}\frac{a}{bk}, \qquad \rho_c = \frac1{3b}.
\end{equation}
Evaluating the curvature scalars corresponding to the Ruppeiner-$V$ and $N$ 
metrics using Eqs.~\eqref{eq:gV-defn},~\eqref{eq:gN-defn}, and~\eqref{eq:thermocurvature-RY},
and then normalizing the temperature and number density to their values at
the critical point, so that $t := T/T_c$ and $\varrho := \rho/\rho_c$, we obtain
\begin{align}
	R_V &= -\frac bV\frac{(\varrho-3)}{3\bkt{\varrho(\varrho-3)^2-4t}^2} \nonumber \\[5pt]
	    &\quad\,\,{}\times \bkt{16t^2 + 2(\varrho-3)\bkt{\varrho(\varrho-6)-3}t
    +3\varrho(\varrho-3)^3},\vphantom{\bigg)} \\[5pt]
	R_N &= \frac{\varrho(\varrho-3)^2}{3N\bkt{\varrho(\varrho - 3)^2
	- 4t}^2}\bkt{\varrho(\varrho-3)^2 - 8t}.
\end{align}
Due to the factor of $\varrho - 3$ in the numerators of both expressions, both
curvature scalars vanish when $V = Nb$. It was argued in~\cite{wei2019ruppeiner}
that this is consistent with Ruppeiner's conjecture,
since this corresponds to the smallest allowable density $b$,
and then the system is effectively a solid of noninteracting particles.
One can also see that both curvature scalars diverge along the spinodal curve for
the van der Waals fluid,
\begin{equation}
	t_\text{sp} = \frac{\varrho(\varrho - 3)^2}{4}.
\end{equation}
However, the two are qualitiatively different otherwise. One can see that $R_N$ 
vanishes and changes sign along the curve
\begin{equation}
	t = \frac{\varrho(\varrho - 3)^2}{8} \equiv \frac{t_\text{sp}}{2},
\end{equation}
whereas $R_V$ is zero and changes sign at
\begin{equation}
	t = \frac1{16}\paren{-9 - 15\varrho + 9\varrho^2 - \varrho^3
+ A\,}\vphantom{\Bigg)},
\end{equation}
where
\begin{equation}
A := \sqrt{(\varrho-3)^2(\varrho^4 - 12\varrho^3 -18\varrho^2 + 180\varrho +9)}.
\end{equation}
These define distinct curves; the results are summarized in Figure~\ref{fig:vdw-rcurves}.
A similar analysis was carried out in Ref.~\cite{wei2019ruppeiner} for the
Ruppeiner-$N$ metric, with the conclusion that since the sign-changing curve lies
within the coexistence region, the van der Waals equation of state must be invalid
there. Then since the curvature scalar never changes sign in the physical region,
interactions in the van der Waals fluid must be uniformly repulsive.
Figure~\ref{fig:vdw-rcurves} emphasizes the need to specify which induced metric
is being used when making reference to Ruppeiner's conjecture. We will provide
further examples shortly.

The ideal gas can be obtained from the van der Waals model by letting the effects of attractive interactions and volume exclusion go to zero, which corresponds to taking the limit $a,b \to 0$ of Eq.~\eqref{eq:Fvdw}.
In this case the critical point disappears, and the Helmholtz free energy becomes
\begin{equation}
F = NkT\paren*[\bigg]{\ln\rho - \frac32\ln T - \frac32\ln\eta}.
\end{equation}
It is straightforward to show, using Eqs.~\eqref{eq:gV-defn} and~\eqref{eq:gN-defn}, that the corresponding Ruppeiner metrics are
\begin{align}
    g_V &= \frac{3N}{2T^2}\,dT^2 + \frac1N\, dN^2,\\[5pt]
    g_N &= \frac{3N}{2T^2}\,dT^2 + \frac N{V^2}\, dV^2.
\end{align}
Both induced metrics for the ideal gas are known to be flat \cite{nulton1985geometry}. This can be shown for the Ruppeiner-$V$ metric by performing the
transformation $r = 2\sqrt N$, $\theta = \sqrt{3/8}\,\ln T$, which yields $g_V = dr^2 + r^2 \,d\theta^2$.
For the Ruppeiner-$N$ metric, the coordinate transformation $x = \sqrt{3N/2}\,\ln T$, $y = \sqrt N \ln V$ yields
$g_N = dx^2 + dy^2$. It follows that both curvature scalars
vanish, $R_V = R_N = 0$.

\section{Vanishing thermodynamic curvature}
\label{sec:vanishing-curvature}

The preceding discussion demonstrates that $R=0$ for the ideal gas, which is why zero curvature is commonly interpreted as the absence of interparticle interactions. However, there is no fundamental reason why the ideal gas should be the only system with this property. If it were, this would reinforce the interpretation of thermodynamic curvature as a direct measure of interparticle interactions. Yet, as we will show, this is not the case---there are other solutions where 
$R$=0 that do not correspond to ideal gases. This raises important questions about the physical nature of these systems, which we will explore in detail. Specifically, we also aim to determine whether the $R=0$ solutions we uncover represent gases that can exist in nature, or if they are merely extraneous, unphysical solutions that should be dismissed as counterexamples to Ruppeiner's conjecture.

We note that our approach to finding flat metrics applies only when two
variables are allowed to vary. In two dimensions, the symmetries of the Riemann curvature
tensor imply that there is only one independent component, which can be shown
is proportional to the curvature scalar $R$. If more than two variables
can fluctuate, then one must show that all components of the Riemann curvature
tensor vanish, or put the metric in manifestly flat form.
Compared to the two-dimensional case we focus on, there are fewer studies of Ruppeiner geometry in three dimensions. See, for example, Refs.~\cite{ruppeiner2020thermodynamic,sahay2017restricted,ruppeiner2007stability}. In practice, however, one may still test the conjecture~\eqref{eq:Rsimxi} and its consequences; the only difference is whether $R = 0$ can be identified with a flat metric.

\subsection{Exact solutions}
\label{sec:exact-solutions}

We begin by finding exact solutions to the equation $R = 0$. We propose
two approaches: the first involves working
in terms of metric components, and the second works directly in terms
of thermodynamic response functions.
The first approach generalizes a result by
Garcia-Ariza \emph{et al.}~\cite{garcia2014geometric} and
provides a clean description of flat Ruppeiner metrics, but
is of limited applicability to the Ruppeiner-$V$ metric.
The second approach, based on response functions,
is partly motivated by the need to provide a better
description of systems with flat Ruppeiner-$V$ metrics. Using both approaches, we
demonstrate the existence of non-ideal gases with either $R_V = 0$ or $R_N = 0$,
which shows that the vanishing of either one of these curvature scalars is not a unique
property of the ideal gas. Additionally, we find that the form of our solutions
here are not readily suited to physical interpretation; we are unable to say whether
the systems they describe exhibit primarily repulsive or attractive interparticle
interactions, or neither. This limitation will be addressed in
Sections~\ref{sec:solutions-from-virial} and~\ref{sec:interpretingB2}.

\subsubsection{Via metric components}

We will construct a class of thermodynamic systems with vanishing curvature scalar
by specifying the components of the Ruppeiner metric. We work in
the Helmholtz free energy representation, so that the Ruppeiner metric is diagonal
and can be written in the form of Eq.~\eqref{eq:ruppmetric-Frep}.
We now outline our construction. Suppose that $g_{TT}$ contains no dependence on $X$,
so that $\partial g_{TT} / \partial X = 0$, and Eq.~\eqref{eq:thermocurvature-RY}
reduces to
\begin{equation}
	R_{Y} = 
	\frac1{\sqrt{g_{TT}g_{XX}}} \pfr{}T\paren{
		\frac1{\sqrt{g_{TT}g_{XX}}}\pfr{g_{XX}}T
	}.
\end{equation}
The expression in parentheses can be rewritten as
\begin{equation}
	\frac1{\sqrt{g_{TT}g_{XX}}} \pfr{g_{XX}}T
	= \frac2{\sqrt{g_{TT}}} \pfr{}T \sqrt{g_{XX}},
\end{equation}
so that if for some free function $A$ we have
\begin{equation}
	\frac1{\sqrt{g_{TT}}}\pfr{}T\sqrt{g_{XX}} = A(X),
	\label{eq:gXX-intermediate}
\end{equation}
then the curvature scalar $R_Y$ vanishes. After integrating both sides
of Eq.~\eqref{eq:gXX-intermediate} with respect to $X$ and introducing another
free function $B$, we obtain
 \begin{equation}
	 g_{XX} = \paren{A(X) \int dT\, \sqrt{g_{TT}} + B(X)}^2.
	 \label{eq:gXX-flatcondition}
\end{equation}
Therefore, if $\partial g_{TT}/\partial X = 0$, then the
metric~\eqref{eq:ruppmetric-Frep} will be flat if $g_{XX}$ is of the form
\eqref{eq:gXX-flatcondition}. For more direct physical interpretation, this
result may be restated in terms of response functions, similar to the view taken in
Eqs.~\eqref{eq:gV-respfxns} and \eqref{eq:gN-respfxns}. This
leads to the following pair of results: the Ruppeiner-$N$ metric is flat
for any system with $\partial C_V / \partial V = 0$ and
\begin{equation}
	B_T = T\paren{A(V) + B(V) \int dT\, \frac{\sqrt{C_V}}{T}}^2,
	\label{eq:BT-flat-comps}
\end{equation}
where $A$ and $B$ are free functions of $V$. Meanwhile, the Ruppeiner-$V$ metric
is flat for any system with $\partial C_V / \partial N = 0$ and
\begin{equation}
	\paren{\pfr {\mu}N}_{T,V} = T\paren{
		A(N) + B(N) \int dT\, \frac{\sqrt{C_V}}{T}
	}^2,
	\label{eq:partialmu-ga}
\end{equation}
where $A$ and $B$ are free functions of $N$. Note that the converse does not generally
hold for both statements. These findings generalize the result
of~\cite{garcia2014geometric} by allowing a wider range of temperature dependence,
and imply the existence of a broad class of metrics
with vanishing thermodynamic curvature.

The result~\eqref{eq:BT-flat-comps} may help explain why the ideal gas has
a flat Ruppeiner-$N$ metric: a direct computation of the response functions for
the ideal gas shows that
\begin{equation}
	C_V = \tfrac32 Nk,\qquad B_T = \frac{NkT}V.
\end{equation}
Since $\partial C_V / \partial V = 0$ and $B_T$ is of the form~\eqref{eq:BT-flat-comps}
with $B(V) = 0$, we conclude that $R_N = 0$. However, the
analogous result~\eqref{eq:partialmu-ga} for the Ruppeiner-$V$ metric is not
applicable to the ideal gas, because  $\partial C_V / \partial N = \tfrac32 Nk \ne 0$
for this system. Therefore, the flatness of the Ruppeiner-$V$ metric must
be due to some other reason. Indeed, the constraint
$\partial C_V / \partial N = 0$ required by Eq.~\eqref{eq:partialmu-ga} seems
generally inapplicable to physical gases, for it requires the heat capacity
to have no dependence on the particle number $N$.
Thus, while Eq.~\eqref{eq:partialmu-ga} describes a system with $R_V = 0$,
it is unclear whether this is a solution of physical significance.

\phantom{x}
\subsubsection{Via thermodynamic response functions}

One way to bypass the difficulty of the previous section is to express
Eq.~\eqref{eq:thermocurvature-RY} directly in terms of thermodynamic
response functions. A similar view was taken in Ref.~\cite{branka2018thermodynamic} to
facilitate the calculation of $R_V$ from molecular dynamics simulation data, where
the full form of the Helmholtz free energy is generally unknown, but response function
data is more readily available. We extended their calculation to include both
induced metrics, with the result
\begin{widetext}
\begin{align}
	R_V &= \frac{1}{NM^*C^*}\bigg\{ T \,\pfr{M^*}T \bkt{1 - \frac12 \hspace{1pt}T
			\hspace{1pt}
	\pfr{\ln(M^*C^*)}{T}} + T^2 \pfr[2]{M^*}T \nonumber\\[10pt]
	    &\hspace{125pt}+ N \pfr{C^*}N\bkt{1 - \frac12 \hspace{1pt}N
		    \hspace{1pt}
    \pfr{\ln(M^*C^*)}{N}} + N^2 \pfr[2]{C^*}N
- \frac12\hspace{1pt}N \hspace{1pt}C^* \pfr{\ln(M^*/C^*)}{N} \bigg\},
\label{eq:RV-respfxn}\\[15pt]
	R_N &= \frac{1}{NB^*C^*}\bigg\{ T \,\pfr{B^*}T \bkt{1 - \frac12 \hspace{1pt}T
			\hspace{1pt}
	\pfr{\ln(B^*C^*)}{T}} + T^2 \pfr[2]{B^*}T
+ V \pfr{C^*}V\bkt{1 - \frac12 \hspace{1pt}V
		    \hspace{1pt}
    \pfr{\ln(B^*C^*)}{V}} + V^2 \pfr[2]{C^*}V \bigg\}.
\label{eq:RN-respfxn}
\end{align}
\end{widetext}
The starred quantities have been nondimensionalized
and normalized to their values for the ideal gas, so that
$C^* := C_V/Nk$, $B^* := B_T/(N k T/V)$, and $M^* := (\partial\mu/\partial N)_{T,V}
/(T/N)$. Note that $B^*$, $C^*$, and $M^*$ are constant
for the ideal gas by definition, so that all the derivatives of these quantities
vanish and $R_V = R_N = 0$  for this system, as expected.
Equations~\eqref{eq:RV-respfxn} and~\eqref{eq:RN-respfxn} provide the conditions
for the curvature scalars $R_V$ and $R_N$ of a thermodynamic system to be zero: the response
functions $C_V$, $B_T$, and $(\partial\mu/\partial N)_{T,V}$
should be such that the right-hand side of these equations
are zero. While a complete characterization of all such response functions seems
out of reach at present, we can find solutions in a few special cases, as we
demonstrate now. Note, in particular, that in what follows, we have no need to demand
$\partial C_V / \partial N = 0$. This bypasses our main difficulty with the use
of Eq.~\eqref{eq:partialmu-ga}.

For ease of analysis, we will consider the class of thermodynamic systems with isochoric
heat capacity of the form $C_V = Nkf(T)$, where $f(T)$ is some positive definite function.
Upon substitution into Eq.~\eqref{eq:RV-respfxn}, the derivatives of $C^*$ with
respect to density vanish, and then $R_V = 0$ if and only if
\begin{align}
	0 &= T\hspace{1pt}\pfr{M^*}T \bkt{1 - \frac12\hspace{1pt}T
	\hspace{1pt} \pfr{\ln(M^* f)}T} \nonumber\\[5pt]
	  &\hspace{80pt}{}+ T^2 \pfr[2]{M^*}T
	- \frac12\hspace{1pt}N f\hspace{1pt}\pfr{\ln M^*}N.
	\label{eq:M-flat-respfxns}
\end{align}
Assuming the heat capacity (hence the function $f$) is known, this provides an
equation that $(\partial\mu/\partial N)_{T,V}$ must satisfy. We will discuss three
simple assumptions that can be made about the form of $M^*$, and in each case
obtain a solution to Eq.~\eqref{eq:M-flat-respfxns}.
\begin{enumerate}[leftmargin=*, label=(\arabic*)]
\item Suppose $M^* = M^*(T)$. Then Eq.~\eqref{eq:M-flat-respfxns} has solution
\begin{equation}
	M^*(T) = c_2 \exp\bkt{ \int_1^T
	\frac{f(y)^{1/2}\,dy}{y(c_1 + \frac12\int_1^y x^{-1}f(x)^{1/2}dx)}},
\end{equation}
where $c_1$ and $c_2$ are constants of integration.
For the special case $f(T) = \text{constant}$, which includes the ideal gas,
this evaluates to
\begin{equation}
M^*(T) = c_2(c_1 + \ln T)^2.
\label{eq:M-Tonly}
\end{equation}
Interestingly, this is disjoint from the ideal gas solution, which has
$M^*(T)$ constant.
\item  Suppose $M^* = M^*(N)$. Then Eq.~\eqref{eq:M-flat-respfxns}
	becomes
	\begin{equation}
		0 = \frac12\hspace{1pt}N\hspace{1pt} f(T)\dfr{\ln M^*}N,
	\end{equation}
	and the only solution is $M^*(N) = \text{constant}$.
\item Finally, suppose that
	$M^*(T,N) = \Theta(T)\mathcal{N}(N)$. Then Eq.~\eqref{eq:M-flat-respfxns}
	can be recast in the form
\begin{align}
	&\frac1f\set*{T \hspace{1pt}\dfr\Theta T\bkt{1 -
		\frac12\hspace{1pt}T\paren{\frac{1}{\Theta}\dfr\Theta T +
\frac{1}{f}\hspace{1pt}\dfr fT}}
+ T^2\dfr[2]\Theta T}\nonumber \\[5pt] &\hspace{80pt}{}
= \frac12\hspace{1pt}N\hspace{1pt}\frac1{\mathcal{N}^2}\dfr{\mathcal{N}}N.
\end{align}
Since the expressions on the left and right-hand sides depend
only on $T$ and $N$, respectively, we may equate both to a separation constant $C$.
First suppose that $C \ne 0$.
Then the equation for the particle number dependence has solution
\begin{equation}
	\mathcal{N}(N) = -\frac{1}{c_1 + 2A \ln N},
\end{equation}
where $c_1$ is a constant of integration. Meanwhile, we were unable to find a general
solution to the equation for the temperature dependence for arbitrary $f$,
so as before, we consider the special case of constant $f$. But
the solution in this case is complex-valued,  so we discard it as unphysical.

We next assume that $C = 0$, and again consider the special case of constant $f$.
Then for the particle number dependence we obtain the solution
$\mathcal{N}(N) = \text{constant}$,
and for the temperature dependence we get $\Theta(T) = c_2(c_1 + \ln T)^2$,
where $c_1$ and $c_2$ are integration constants. This
merely recovers the solution~\eqref{eq:M-Tonly} from earlier.
\end{enumerate}
To summarize briefly, assumptions (1)--(3) have led to a novel
solution to $R_V = 0$, namely $C_V \sim Nk$ and  $(\partial\mu/\partial N)_{T,V}$
described by Eq.~\eqref{eq:M-Tonly}. It is interesting to note that this
response function diverges to positive infinity as $T \to 0$, unlike
the ideal gas, which has $(\partial\mu/\partial N)_{T,V} \to 0$ in
the same limit. We expect that further
solutions can be obtained by investigating more general functions $f$; in this
discussion we have restricted ourselves to cases where $f$ is constant. For
our purposes, it suffices to note that exact, non-ideal gas solutions to $R_V = 0$
exist, and that this claim can be made without adding the assumption
that $\partial C_V/ \partial N = 0$, as required by Eq.~\eqref{eq:partialmu-ga}.

This concludes our brief analysis of Eq.~\eqref{eq:M-flat-respfxns}.
Let us once again stress the importance of this expression: it provides a condition
on $(\partial\mu/\partial N)_{T,V}$ that any thermodynamic system with $C_V = Nkf(T)$
must satisfy in order for the associated curvature scalar $R_V$ to be zero.
In this regard, Eq.~\eqref{eq:M-flat-respfxns}
is a special case of Eq.~\eqref{eq:RV-respfxn}, which can be applied to all
thermodynamic systems, regardless of the form of the isothermal
heat capacity. However, our assumption about the form of $C_V$ does not seem
too restrictive, and is a good starting-point for finding
non-ideal gas systems with vanishing $R_V$.

Similar considerations may be applied to Eq.~\eqref{eq:RN-respfxn} to find systems
with vanishing $R_N$; we shall not do so here.

\subsection{Solutions from the virial expansion of gases}
\label{sec:solutions-from-virial}

An approach which is more naturally suited for the study of interparticle interactions,
and which we shall adopt for the rest of this paper, is to work with
the virial expansion of the Helmholtz free energy~\cite{mayer1940statistical}
\begin{align}
	F &= N k T \Bigl( \ln \rho - \tfrac32 \ln(\gamma T) - 1\nonumber\\[3pt]
		  &\hspace{4em}{}+ B_2(T)\hspace{1pt} \rho +
	  \tfrac12 B_3(T)\hspace{1pt}\rho^2 + \cdots \Bigr) \nonumber \\[5pt]
		  &= N k T \Bigl( \ln \paren{\frac NV} - \tfrac32 \ln(\gamma T)
			  - 1\nonumber\\[3pt]
		  &\hspace{4em}{}+ B_2(T)\hspace{1pt} \frac NV +
	  \tfrac12 B_3(T) \paren{\frac NV}^2 + \cdots\Bigr),
\label{eq:F_virialexp}
\end{align}
where $\gamma$ is a constant and $\rho \equiv N/V $ is the number density of the gas.
This expansion is valid in the low density, high temperature regime. The virial
coefficients $B_2,B_3,\dots$ are functions of temperature alone, and can be
obtained by evaluating integrals of the intermolecular potential. In particular,
then, a given potential specifies a set of virial coefficients. Since  $F = F(T,V,N)$
is a thermodynamic potential, specifying all of the virial coefficients amounts to
giving a complete thermodynamic description of the system, with the first few
coefficients providing the leading-order corrections to the ideal gas.

The virial expansion~\eqref{eq:F_virialexp}
fully specifies the functional dependence of the Helmholtz free energy
$F$ on $N$ and~$V$, leaving only the virial coefficients to be determined.
Using Eqs.~\eqref{eq:gV-defn},~\eqref{eq:gN-defn}, and~\eqref{eq:thermocurvature-RY}
to compute the curvature scalar for the Ruppeiner-$V$ and Ruppeiner-$N$ metrics,
and then expanding in powers of $\rho$,
we obtain
\begin{widetext}
\begin{align}
R_V &=
\frac{1}{3V} \paren{T^2 \dfr[2]{B_2}{T}-2 T \dfr{B_2}T-3 B_2}\nonumber\\[10pt]
  &\qquad {}+\frac{1}{9V}
  \Biggl(-20 T^2 \paren{\dfr{B_2}T}^{\!2}+24 T \dfr{B_2}T B_2+4 T^4
	  \dfr[3]{B_2}T \dfr{B_2}T
	  + 8 T^3 \,\dfr[2]{B_2}T\dfr{B_2}T \label{eq:RVorders} \\[5pt]
	    &\qquad\qquad\qquad{}+36
  B_2{\hspace{0.1pt}^2}+3 T^2\, \dfr[2]{B_3}T
-12 T \,\dfr{B_3}T-27 B_3\Biggr)\rho + O(\rho^2), \nonumber\\[15pt]
R_N &=
\frac1{3V}\paren{2T^2\dfr[2]{B_2}{T}}\nonumber\\[10pt]
    &\qquad {}+\frac2{9V}
  \Biggl(
  2 T^4 \dfr[3]{B_2}{T} \dfr{B_2}T+T^4 \left(\dfr[2]{B_2}T\right)^2
  +8 T^3 \dfr{B_2}T \dfr[2]{B_2}T-6 T^2 \left(\dfr{B_2}T\right)^2\label{eq:RNorders}\\[5pt]
  &\qquad\qquad\qquad{}-3 T^2 B_2 \dfr[2]{B_2}T+3 T^2 \dfr[2]{B_3}T+6 T B_2 \dfr{B_2}T
	  -3 T \dfr{B_3}T
  \Biggr)\rho + O(\rho^2). \nonumber
\end{align}
\end{widetext}
Rather than search for an exact solution in the manner of
Section~\ref{sec:exact-solutions}, this suggests a different approach to finding a flat
Ruppeiner metric, namely, the curvature scalar $R$ will be zero if every coefficient of
$\rho$ vanishes. With this in mind, we highlight an important feature of
Eqs.~\eqref{eq:RVorders} and \eqref{eq:RNorders}.
In both expressions, the $O(1)$ coefficient involves only $B_2$,
and the $O(\rho)$ coefficient introduces $B_3$. This pattern continues up to higher orders,
with each new order in $\rho$ introducing an additional virial coefficient.

Now we demand that the curvature scalars $R_V $ and $R_N$ vanish to leading order
in $\rho$. To do this, we equate the $O(1)$ coefficients
of $\rho$ in both Eqs.~\eqref{eq:RVorders} and \eqref{eq:RNorders} to zero,
then solve the resulting equations for the second virial coefficient.
In both cases, we obtain a linear second
order ordinary differential equation for $B_2$.
If we demand that $R_V$ vanishes to $O(1)$ in $\rho$, we get
\begin{equation}
	B_2(T) = c_1 \, T^{(3-\sqrt{21})/2} + c_2 \, T^{(3+\sqrt{21})/2},
      \label{eq:B2-V-twoterm}
\end{equation}
where $c_1$ and $c_2$ are constants of integration, while for $R_N$ we obtain
\begin{equation}
	B_2(T) = c_3 + c_4\, T,
      \label{eq:B2-N-twoterm}
\end{equation}
where $c_3$ and $c_4$ are constants of integration.
We now ask whether it is possible to select virial coefficients
to make the curvature scalars $R_V$ and $R_N$
vanish to all orders of~$\rho$. We note from our discussion in
Section~\ref{sec:exact-solutions} that one scalar can vanish while the other
does not. Let us begin by considering the curvature scalar $R_V$.
Starting from Eq.~\eqref{eq:B2-V-twoterm} and choosing $c_1 = \alpha > 0$
and $c_2 = 0$, we have that $R_V$ vanishes to $O(1)$ if
\begin{equation}
	B_2(T) = \alpha\, T^{(3-\sqrt{21})/2}.
	\label{eq:B2-V-flat}
\end{equation}
Then we substitute this into the $O(\rho)$ coefficient in Eq.~\eqref{eq:RVorders}
and solve for $B_3$ to get
\begin{align}
B_3(T) &= \tfrac43(-6 + \sqrt{21})\, \alpha^2\, T^{3-\sqrt{21}}\nonumber\\[5pt]
       &\qquad\qquad\qquad{}+ c_3\, T^{(5-\sqrt{61})/2} + c_4\, T^{(5+\sqrt{61})/2}.
\label{eq:B3-V-flat}
\end{align}
We then set both integration constants $c_3$ and $c_4$ to zero, leaving us with
a third virial coefficient $B_3(T) \sim T^{3-\sqrt{21}}$. This procedure can
be iterated as follows: we insert all previously determined virial coefficients
$B_2,B_3,\dots,B_2$ into Eq.~\eqref{eq:RVorders}, so that all terms until the
$O(\rho^{n-2})$ coefficient vanish, by construction. Then we set the
$O(\rho^{n-1})$ coefficient equal to zero, and solve the resulting ordinary
differential equation for $B_{n+1}$. Finally, we set any unspecified integration
constants in this solution to zero. We found no obstacle, barring computational
time, to obtaining higher virial coefficients this way, and our code was
capable of generating coefficients up to $B_{15}$ in around one hour of runtime.
Indeed, in examining the higher-order terms of Eq.~\eqref{eq:RVorders}, we found
that setting the $O(\rho^{n})$ coefficient to equal to zero amounted to writing
a linear second-order ordinary differential equation for the virial coefficient,
provided that $B_2,B_3,\dots,B_{n+1}$ were already specified. Since solutions
to these equations always exist, we conjecture that, using this construction, it
is possible to choose virial coefficients such that the thermodynamic curvature
scalar $R_V$ vanishes to all orders in $\rho$.

We found that all the virial coefficients obtained via this construction were
of the form
\begin{equation}
	B_n(T) = \overline{B}_n\,\alpha^{n-1}T^{(3-\sqrt{21})(n-1)/2},
	\label{eq:Bn-V-flat}
\end{equation}
where $\overline B_n$ is a constant.
For example,
in the virial coefficients~\eqref{eq:B2-V-flat} and \eqref{eq:B3-V-flat}
already determined, we have $\overline B_2 = 1$ and $\overline B_3 = \tfrac43
(-6+\sqrt{21}) \approx -1.8899$. With the provision~\eqref{eq:Bn-V-flat},
specifying the numbers $\overline{B}_n$ amounts to specifying the virial
coefficients themselves. For reference, the first 15 of these values are collected
in Table~\ref{table:reducedvirials}. Note that the virial coefficients oscillate
between negative and positive values, and that they grow in magnitude by a factor
of around 10 with each order. This suggests that the virial series obtained
by this construction converges slowly, if at all.

\begin{table}[]
\renewcommand{\arraystretch}{1.3}
\caption{Numerical prefactors for virial coefficients yielding a thermodynamic curvature scalar
	$R_V$ which vanishes to $O(\rho^{13})$. These oscillate between positive
	and negative values, and grow in magnitude
	by a factor of around 10 with each order.}
\label{table:reducedvirials}
\vspace{1em}

\begin{tabularx}{0.8\columnwidth}{XX}
	\toprule
	$n$ & $\overline B_n$  \\
\midrule
	2   & 1 \\
3   & $-1.8899$ \\
4   & $7.24728$  \\
5   & $-36.5603$ \\
6   & $214.284$  \\
7   & $-1381.61$ \\
8   & $9518.08$  \\
9   & $-68846.3$ \\
10  & $516959$  \\
11   & $-3.998\times 10^6$  \\
12   & $3.168\times 10^7$ \\
13   & $-2.5606\times 10^8$  \\
14   & $2.1046\times 10^9$ \\
15  & $-1.7547\times 10^{10}$  \\
\bottomrule
\end{tabularx}
\end{table}

Curiously, the temperature dependence specified by Eq.~\eqref{eq:Bn-V-flat} is
identical to that of the virial coefficients for the inverse-power law
potential~\cite{wheatley2005inverse},
which is defined by the intermolecular potential $u(r) = \gamma/r^{n}$ and
$n \ge 3$. This agreement is not entirely coincidental, for in
Section~\ref{sec:interpretingB2} we will show that the second virial coefficient
of this construction corresponds to the choice of an inverse power-law
potential with $n = (3+\sqrt{21})/2 \approx 3.7913$. This may also clarify the
slow convergence in the virial expansion, since it has previously been observed that
the virial series for these potentials is poorly behaved close to the thermodynamically
allowed limit of $n \to 3+$ \cite{wheatley2005inverse,barlow2012asymptotically}.

We now turn our attention to $R_N$, and ask whether we can choose
virial coefficients such that the expansion~\eqref{eq:RNorders} yields zero
to all orders in $\rho$. The reasoning is similar to what was used for the
other curvature scalar.
Choosing $c_4 = 0$ in Eq.~\eqref{eq:B2-N-twoterm},
we have that $R_N$ vanishes to $O(1)$ if
\begin{equation}
B_2(T) = c_3,
\end{equation}
where $c_3$ is some constant. Putting this back into Eq.~\eqref{eq:RNorders}
and solving for $B_3$ yields
\begin{equation}
B_3(T) = c_5\, T^2 + c_6,
\end{equation}
where $c_5$ and $c_6$ are integration constants. We then set $c_5 = 0$, leaving
us with $B_3 \sim \text{constant}$. If we continue this process iteratively, we
find that for every coefficient in Eq.~\eqref{eq:RNorders}, the choice
\begin{equation}
B_n(T) = \overline B_n,
\label{eq:Bn-N-flat}
\end{equation}
where $\overline B_n$ is some constant, suffices to make $R_N$ vanish up to
arbitrarily large order in $\rho$. The result that a set of constant virial coefficients
comprises a solution to $R_N = 0$ can be explained by the use
of Eq.~\eqref{eq:gXX-flatcondition}. To see
how, let us once again write the Helmholtz free energy in terms of the virial
expansion, inserting Eq.~\eqref{eq:Bn-N-flat} for the virial coefficients. This gives
\begin{align}
F &= NkT \bigg[ \ln\paren{\frac NV} - \tfrac32 \ln(\gamma T)- 1
	+ \sum_{i=2}^\infty \overline B_i \paren{\frac NV}^{i-1}\,
\bigg].
\end{align}
Direct calculation of metric components via Eq.~\eqref{eq:gN-defn} gives
\begin{align}
	g_{TT} &= \frac{3Nk}{2T^2}, \\[5pt]
	g_{VV} &= \frac{Nk}{V^2} + \sum_{i=2}^\infty \frac{i \overline B_i
	N^i k}{V^{i+1}}.
\end{align}
Since $\partial g_{TT} / \partial V =0$ and $g_{VV}$ takes the
form of Eq.~\eqref{eq:gXX-flatcondition} with $A(V) = 0$, it follows that $R_N = 0$ 
for this system, as we wanted to show.

Interestingly, this provides another example of a system such that $R_N$ vanishes
everywhere, but $R_V$ is nonzero. If we proceed to calculate the other
scalar curvature $R_V$ associated with~\eqref{eq:Bn-N-flat}, we obtain
\begin{equation}
R_V
= -\frac{\vphantom{\big(}
\overline B_2 + 3\rho\hspace{1pt} \overline B_3 + 6\rho^2\hspace{1pt} \overline B_4
+ 10 \rho^3\hspace{1pt} \overline B_5 + \cdots}
{V\paren*[\big]{1 + 2\rho\hspace{1pt} \overline B_2 + 3\rho^2\hspace{1pt} \overline B_3
	+ 4\rho^3\hspace{1pt} \overline B_4
+ 5\rho^4\hspace{1pt} \overline B_5 + \cdots}^2}.
\label{eq:RV-hardspherelike}
\end{equation}
The implications of this are twofold. First, we note that this result implies
$R_V \ne 0$ everywhere, provided that $\overline B_i > 0$ for each $i$. Thus,
the system described by the virial coefficients~\eqref{eq:Bn-N-flat} provides
yet another example of a system which is flat with respect to one, but not both,
of the induced metrics $g_V$ and $g_N$. Second, the virial coefficients~\eqref{eq:Bn-N-flat} are similar to those of the hard-sphere gas, which is not
a non-interacting system, and is in fact repulsive everywhere. Thus, the result
$R_V < 0$, which suggests attractive interactions,
is inconsistent with Ruppeiner's conjecture. A similar result was obtained
in~\cite{ruppeiner2021thermodynamic}, where it was found that $R_V < 0$ at
all temperatures and densities for gases modeled by the Carnahan-Starling equation
of state. It has also been noted that the hard-sphere limit of the inverse-power
potential results in a scalar curvature that is negative
everywhere~\cite{branka2018thermodynamic}. The authors of
Ref.~\cite{ruppeiner2021thermodynamic} argued, then, that a negative scalar curvature
seems to be a persistent feature of
hard-sphere gases, in seeming contradiction to the predictions of thermodynamic
geometry. The result~\eqref{eq:RV-hardspherelike} shows that this is
a generic feature of \emph{any} gas with constant, positive virial coefficients,
of which hard-sphere gases are only a particular example.

\subsection{An extension to Ruppeiner's conjecture}

Having now seen examples where one of the curvature scalars for a system is zero,
we are led to ask whether it is possible to make both scalars
vanish, i.e., whether there exist systems such that $R_V = R_N = 0$.
We know this to hold for the ideal gas; we now demonstrate that the ideal gas is the only system for which both curvature scalars vanish to all orders
in $\rho$. First, by comparing Eqs.~\eqref{eq:B2-V-twoterm} and \eqref{eq:B2-N-twoterm}, we see that these virial coefficients are generally
distinct unless $c_1 = c_2 = c_3 = c_4 = 0$. This choice of constants corresponds to $B_2 = 0$.
Thus, if we wish for both curvature scalars vanish to $O(1)$, then we
must choose  $B_2 = 0$. We substitute this virial coefficient into Eqs.~\eqref{eq:RVorders}
and \eqref{eq:RNorders}, and then solve for the third virial coefficient $B_3$
which makes the curvature scalar vanish to $O(\rho)$.
For $R_V$, the solution is
\begin{equation}
	B_3(T) = c_1\, T^{(5-\sqrt{61})/2} + c_2\,T^{(5+\sqrt{61})/2},
\end{equation}
while for $R_N$, the solution is
\begin{equation}
B_3(T) = c_3\, T^2 + c_4.
\end{equation}
Clearly, these only coincide when $c_1 = c_2 = c_3 = c_4 = 0$, so if we require that both $R_V$ and $R_N$ vanish to $O(\rho)$, then we must choose
$B_2 = B_3 = 0$.

A similar pattern continues to higher orders, which yields
the following result:
For any given $n \ge 0$, the unique solution
to $R_V = R_N = 0$, up to $O(\rho^{n})$, is to set $B_2 = B_3 = \cdots = B_{n+2} = 0$.
Since $n$ can be made arbitrarily large, it follows that the
unique solution to $R_V = R_N = 0$ is to set all virial coefficients equal to zero. This amounts to selecting the ideal gas.
For this it suffices to show that if $B_2, B_3, \dots, B_{n-1}$ are all zero, then
the virial coefficients that yield a zero $O(\rho^{n-2})$ coefficient for
the curvature scalars  $R_V$ and $R_N$, respectively, are
\begin{align}
	B_n &= c_1 \, T^{\hspace{0.5pt}a_{n+}}
	+ c_2\, T^{\hspace{0.5pt}a_{n-}},\label{eq:Bn-V-ordern} \\[5pt]
	B_n &= c_3 \, T^{n-1} + c_4, \label{eq:Bn-N-ordern}
\end{align}
where
\begin{equation}
    a_{n\pm} = \frac{(2n-1) \pm \sqrt{10n^2 - 10n + 1}}{2}.
\end{equation}
These virial coefficients can only coincide if $c_1
= c_2 = c_3 = c_4 = 0$, so that  $B_n$ must also be zero, as we wanted to show.
We have verified, up to $O(\rho^{13})$, that the
coefficients given by Eqs.~\eqref{eq:Bn-V-ordern}
and \eqref{eq:Bn-N-ordern} are solutions as described.

The viewpoint which emerged in Section~\ref{sec:exact-solutions} 
was that the ideal gas is not uniquely specified by the vanishing of
either $R_V$ or $R_N$. Thus, although it is certainly privileged in a physical
sense, the ideal gas seemed to be only one of a wide number of systems with vanishing
thermodynamic curvature. Our result in this section, which we obtained by working with the
virial expansion of gases, suggests that the ideal gas remains privileged
in thermodynamic geometry as the only system for which both $R_V = R_N = 0$.
We currently lack a more rigorous proof of this statement, but the work presented
here provides a compelling first step in this direction.
This further underscores the importance of studying the two metrics that
can be induced from the full Ruppeiner metric. While the vanishing of one curvature
scalar is insufficient to conclude the absence of interactions, it appears that the
relationship between thermodynamic curvature and interactions can be preserved,
provided that both curvature scalars are taken into account.
We therefore propose an extension to
Ruppeiner's conjecture: there is some function $f$ such that
\begin{equation}
	\abs{f(R_V,R_N)} \sim \xi^{d}
\end{equation}
and $f(0,0) = 0$. Clearly, there is much left to be determined about the form of $f$;
our main point is that both induced metrics should be taken into account when studying
the Ruppeiner geometry of a system.

In closing, we have used the virial expansion to obtain two main results:
(1) the vanishing of either $R_V$ or $R_N$ alone is insufficient to conclude the
absence of interactions, and
(2) the ideal gas appears to be uniquely characterized in Ruppeiner geometry
by having $R_N = R_V = 0$ to all orders in the number density $\rho$.
This led us to propose an extension of Ruppeiner's conjecture.

\subsection{Intermolecular potentials from the second virial coefficient}
\label{sec:interpretingB2}

Having determined virial coefficients that result in a vanishing curvature scalar, we turn to the question of interpretation. It seems reasonable to us that a nonzero second virial coefficient must correspond to a nontrivial potential, i.e., the presence of interactions. Our work in this section will flesh out this line of thinking.

For sufficiently simple interactions between particles
(i.e.~no angular dependence), the second virial coefficient is given by
\cite{hirschfelder1964molecular}
\begin{equation}
	B_2(\beta) = 2\pi \int_0^\infty r^2\paren{1 - e^{-\beta u(r)}}\, dr,
	\label{B2-defn}
\end{equation}
where $\beta \equiv 1 / kT$ and
where $u(r)$ is the intermolecular potential as a function of separation between molecules.
The approach we shall adopt here is to regard Eq.~\eqref{B2-defn} as an integral equation
for the intermolecular potential $u(r)$. We note that solutions to this equation may not be unique.

\begin{figure}[]
\centering
\includegraphics[width=\columnwidth]{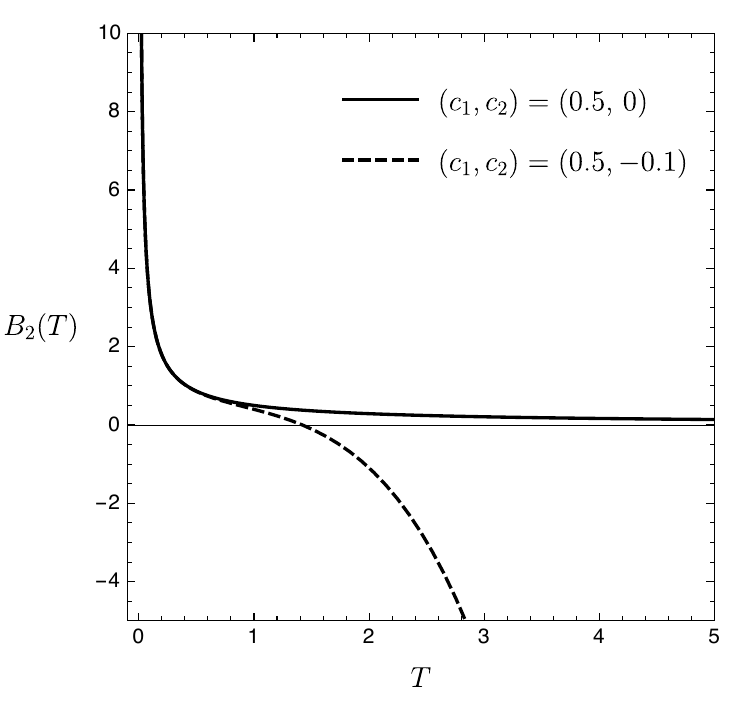}
\caption{Plots of second virial coefficients obtained from
Eq.~\eqref{eq:B2-V-twoterm}; these cause $R_V$ to vanish to $O(1)$ in the number
density $\rho$. When $c_2$ is chosen to be nonzero, the second virial
coefficient diverges to negative infinity as $T \to \infty$.}
\label{fig:b2plots}
\end{figure}

First, we show that the definition of $B_2$ imposes
a few broad constraints on the form of the intermolecular potential.
This will provide better justification for the choice of
integration constants in  Eq.~\eqref{eq:B2-V-twoterm} leading to Eq.~\eqref{eq:B2-V-flat}.
To do this, we observe from Eq.~\eqref{B2-defn} that
\begin{equation}
	\dfr[2]{B_2}{\beta} = - 2\pi \int_0^\infty u(r)^2r^2 e^{-\beta u(r)}\,dr < 0.
\end{equation}
That is, the second derivative of the virial coefficient $B_2(\beta)$ is 
strictly negative, regardless of the form of $u(r)$. It is straightforward to
show that imposing this condition on Eq.~\eqref{eq:B2-V-twoterm} amounts to requiring
that $c_1 > 0$ and $c_2 \le 0$.
We exhibit two characteristic plots in Fig.~\ref{fig:b2plots}; note that
our choice of constants in Eq.~\eqref{eq:B2-V-flat} belongs to the same class
of solutions as the solid curve.

The following result, obtained by Frisch and Helfand, is also useful:
suppose $u(r)$ attains negative values on some interval of $0 < r <\infty$.
Then $B_2(\beta)$ has a maximum if and only if
\begin{equation}
\int_0^\infty u(r) \,r^2\,dr > 0.
\label{frischcondition}
\end{equation}
On the other hand, when $u(r) \ge 0$ for all $0 < r < \infty$, then no maximum in
$B_2(\beta)$ can occur~\cite{frisch1960conditions}. This implies
that if $B_2(\beta)$ has a maximum, then the potential $u(r)$
must attain negative values somewhere and satisfy Eq.~\eqref{frischcondition}.
For the virial coefficient
in Eq.~\eqref{eq:B2-V-twoterm}, which exhibits no maximum for $c_1 > 0$,  $c_2 \le 0$
(compare Fig.~\ref{fig:b2plots}),
we arrive at the following conclusion: the corresponding potential $u(r)$ 
is either strictly positive (i.e.~repulsive), or has an attractive part
that is ``hard'' enough to violate Eq.~\eqref{frischcondition}.

Let us now obtain more detailed information about the potential by solving Eq.~\eqref{B2-defn} for $u(r)$.
The method we describe for inverting the second virial coefficient was studied in Refs.~\cite{keller1959determination,jonah1966direct,maitland1972direct,maitland1985MolPh}; see also the Appendix for a comment on these papers.
Suppose that the pair potential has the limiting behavior
$u(r) \to \infty$ as $r \to 0$ and $u(r) \to 0$ as
$r \to \infty$. Upon integration by parts, Eq.~\eqref{B2-defn} yields
\begin{equation}
B_2(\beta) = -\frac{2\pi\beta}{3}\int_0^\infty r^3 \frac{du}{dr}e^{-\beta u(r)}\,dr.
\label{b2-uderiv}
\end{equation}
Next suppose that $u(r)$ is monotonic. This assumption, which may seem restrictive,
is given further justification in Appendix~\ref{app:inversion}.
By changing integration variable to $u$, we obtain
\begin{equation}
B_2(\beta) = \frac{2\pi\beta}{3}\int_0^\infty r^3
e^{-\beta u}\,du.
\label{eq:monotonicB2}
\end{equation}
The integral on the right defines a Laplace transform, so that we have
\begin{equation}
r^3 = \frac{3}{2\pi}L^{-1}
	\paren{\frac{B_2(\beta)}{\beta}},
	\label{eq:jonahrowlinson}
\end{equation}
where $L^{-1}$ denotes the inverse Laplace transform.
If the inverse transform exists, then this relation can be inverted to obtain an expression for the intermolecular pair potential $u(r)$.

It seems most appropriate, both from a pedagogical perspective and as a
self-consistency check, to demonstrate the inversion procedure for the
second virial coefficient of
a system with a known intermolecular potential $u(r)$. However, these systems are
fairly difficult to come by, and we are further limited by the requirement
that $u(r)$ be monotonic, so that even the van der Waals model cannot be treated
by this procedure.
Therefore, our use of the inversion procedure will be limited to the
second virial coefficients~$B_2$ given in Eqs.~\eqref{eq:B2-V-twoterm}
and~\eqref{eq:B2-N-twoterm}.

We will specialize further to Eq.~\eqref{eq:B2-V-flat},
rather than the more general solution in Eq.~\eqref{eq:B2-V-twoterm}, for
two reasons. First, this result is a particularly interesting
special case, because it is associated with the
procedure, described by Eq.~\eqref{eq:Bn-V-flat}, for
generating virial coefficients yielding a curvature scalar
that vanishes to  high order in~$\rho$.
Second, and more pragmatically, the inversion method fails to work
for the general result, as we will demonstrate below.

We begin with the result in Eq.~\eqref{eq:B2-V-flat}, noting that
\begin{equation}
	\frac{B_2(\beta)}{\beta}
	= \frac{\alpha}{\beta^{(5-\sqrt{21})/2}},
	\label{eq:genericB2}
\end{equation}
so that by using the relation \cite{schiff2013laplace}
\begin{equation}
	L^{-1}\paren{\frac{1}{s^\nu}}
	= \frac{t^{\nu-1}}{\Gamma(\nu)},
	\qquad \nu > 0
	\label{eq:laplace-inversion-invpower}
\end{equation}
in Eq.~\eqref{eq:jonahrowlinson} and then solving for $u(r)$, we obtain
\begin{equation}
u(r) = \frac{\gamma}{r^{(3+\sqrt{21})/2}},
\qquad
\gamma := \paren{\frac{2\pi \Gamma(\frac{5-\sqrt{21}}2)}{3\alpha}}^{\frac{2}
{3-\sqrt{21}}}.
\label{eq:invertedpotential}
\end{equation}
This explains our requirement that $\alpha > 0$ in Eq.~\eqref{eq:B2-V-flat}, for otherwise
 $\gamma$ would be undefined or have an imaginary component.
 Our result here for $u(r)$
shows that the second virial
coefficient in Eq.~\eqref{eq:B2-V-flat}, which causes $R_V$ to vanish to
$O(1)$ in the number density $\rho$, corresponds to a repulsive inverse
power-law potential with $n = (3+\sqrt{21})/2 \approx 3.7913$. To the best of our knowledge,
this result has never been stated explicitly in the literature, although
it is contained implicitly in Eqs.~(33) and (7) of Refs.~\cite{ruppeiner2005riemannian}
and~\cite{branka2018thermodynamic}, respectively. One interpretation of this result,
using Ruppeiner's conjecture, is that the inverse power-law gas with $n = (3 + \sqrt{21})/2$ is somehow privileged by having interparticle interactions that are effectively zero at sufficiently low densities. Such an interpretation, however, would be inconsistent with the results we've
established thus far---namely, that the vanishing of $R_N$ or $R_V$ is generally
insufficient to conclude the absence of interactions.
We also offer a brief comment on the general solution in Eq.~\eqref{eq:B2-V-twoterm}.
While one can still apply Eq.~\eqref{eq:jonahrowlinson} to this solution, the outcome is an expression of the form
$r^3 = a u^{(3-\sqrt{21})/2} + b u^{(3+\sqrt{21})/2}$, where $a$ and $b$ are constants.
We are unable to find solutions to this equation.

Next we consider the virial coefficient of Eq.~\eqref{eq:B2-N-twoterm}, which causes
$R_N$ to vanish to $O(1)$. This is given by
\begin{equation}
\frac{B_2(\beta)}{\beta} = \frac{c_3}{\beta} + \frac{c_4}{\beta^2}.
\end{equation}
The relation~\eqref{eq:laplace-inversion-invpower}, together with
Eq.~\eqref{eq:jonahrowlinson}, implies that this corresponds to the intermolecular
potential
\begin{equation}
u(r) = -\frac{c_3}{c_4} + \frac{2\pi}{3c_4}\, r^3.
\end{equation}
However, this solution must be discarded, because it is incompatible with the assumption that $u(r) \to \infty$ as $r\to 0$ and $u(r) \to 0$ as $r\to \infty$.
It is still possible to interpret
the virial coefficient given in Eq.~\eqref{eq:B2-N-twoterm} by more direct means. Recall that the choice
$c_4 = 0$ leads to a constant second virial coefficient. We claim that this corresponds
to a hard-sphere interaction
\begin{equation}
	u_\text{HS}(r)
	=
	\left\{\ 
\begin{array}{cl}
	\infty &\quad (r < \sigma),\\[10pt]
	0 &\quad (r > \sigma),
\end{array}
	\right.
\end{equation}
which can be shown by a direct computation:
substituting into Eq.~\eqref{B2-defn} gives
\begin{align}
B_2(\beta)
= 2\pi \int_0^\infty r^2\paren{1 - e^{-\beta u_\text{HS}}}dr
&= \frac{2\pi \sigma^3}{3},
\end{align}
which is a positive constant. Thus, the virial
coefficient in Eq.~\eqref{eq:B2-N-twoterm} can be made to correspond to a hard-sphere potential by the choice $c_3 > 0$ and $c_4 = 0$; in particular, the constant  $c_3$ fixes
a diameter $\sigma$ for the spheres.

We mention one last point. Since the second virial coefficient $B_2$
given by Eq.~\eqref{eq:Bn-V-flat} corresponds to an inverse power-law potential with $n = (3+\sqrt{21})/2$, it is natural to ask whether the remaining virial coefficients in Table~\ref{table:reducedvirials} also correspond to this potential. This appears not to be the case. One can show that the ratio $B_3/B_2^2$ is a constant for an inverse-power potential with given~$n$, and the same is true for the virial coefficients defined by Eq.~\eqref{eq:B2-V-flat}. Thus, we can compare the ratio $B_3/B_2^2$ calculated for the $n =(3+\sqrt{21})/2$ potential against the ratio given in Table~\ref{table:reducedvirials}. The $B_3$ integral for the inverse power-law potential can be computed numerically using Fourier transforms, with the result that $B_3/B_2^2 = 0.1183$. On the other hand, the virial coefficients given by Eq.~\eqref{eq:Bn-V-flat}
satisfy $B_3/B_2^2 = -1.8899$. Due to this discrepancy, it would be premature to identify the inverse power-law potential with vanishing $R_V$, as we identified a hard-sphere potential with vanishing $R_N$. Further work is needed in this regard.

\section{Conclusions}
\label{sec:conclusions}

We have studied the existence of flat Ruppeiner metrics, and provided a characterization
of both exact and approximate solutions to $R = 0$. To obtain exact solutions,
we exploited the fact that the Ruppeiner metric is diagonal in the Helmholtz free
energy representation. This allowed us to obtain conditions on the metric components
and thermodynamic response functions of the system
that would ensure the vanishing of the curvature scalar. We found that these
solutions corresponded to non-ideal gases, since the form of the response functions
obtained differed from those of the ideal gas.
However, due to the way in which we obtained our solutions, we were unable to say
anything conclusive about the nature of the interparticle interactions in these
systems.

To obtain approximate solutions, we used the fact that the Helmholtz free energy~$F$
can be written as an expansion in the virial coefficients of the system, which
are in turn specified by integrals of the intermolecular potential.
This expansion fixes the dependence of $F$ on the particle number $N$ and volume $V$,
leaving only the virial coefficients to be specified. By demanding that $R_V$ and
$R_N$ vanish to $O(1)$ in the number density $\rho$, we obtained
a second-order ordinary differential equation in the second virial coefficient $B_2$.
We found that there exists no nontrivial set of virial coefficients $B_2,B_3,\dots,B_n$
such that both $R_V$ and $R_N$ vanish to $O(\rho^{n-2})$.
This comprises one of the key results of this work,
and shows that the vanishing of both curvature scalars is a unique feature
of the ideal gas. Thus, while the vanishing of either $R_V$ or $R_N$ alone
does not characterize the ideal gas---our exact solutions
provide a number of counterexamples---it seems that the ideal gas is privileged
in Ruppeiner geometry by the property $R_V = R_N = 0$. To the best of our knowledge, this
interpretation has not been put forward in the literature.
Then, we carried out an inversion procedure on the second virial coefficients
$B_2$ obtained in this way, and showed that these correspond to nontrivial
intermolecular potentials. In particular, $R_V$ and $R_N$ vanish to $O(1)$ in $\rho$ for an inverse-power potential with $n = (3+\sqrt{21})/2$ and a hard-sphere potential, respectively.
We also outlined two methods to choose virial
coefficients $B_2,B_3,\dots$ such that  $R_V$ or $R_N$ vanishes to
all orders in $\rho$.

Our results more clearly indicate the limitations of Ruppeiner's conjecture: we have provided a number of examples to show that
$R_V = 0$ or $R_N = 0$ is not sufficient to conclude the absence of interactions. Instead, we conjecture that any relationship between interactions and curvature must take both curvature scalars into account.
We hope that these findings will compel others to study the Ruppeiner-$N$ metric, which appears to be another fundamental object in thermodynamic geometry.  It would be interesting to see, for example, whether an interpretation can be given for the sign of $R_N$, similar to how the sign of $R_V$ appears to indicate the nature of interactions. A similar formulation of our conjecture may also be useful in black hole thermodynamics, although there seem to be no clear-cut analogues of the Ruppeiner-$V$ and $N$ metrics in this case. We leave these considerations to future work.

\appendix*

\section{On the inversion of second virial coefficients}
\label[appendix]{app:inversion}

In this appendix, we comment on the Laplace inversion method described by Keller
and Zumino~\cite{keller1959determination}, and address a claim by
Maitland \emph{et al.}~\cite{maitland1985MolPh} that it
is possible to uniquely determine the intermolecular potential from the
second virial coefficient. Our main result is that the inversion method can only
be applied to monotonic potentials, contrary to what is claimed in the literature.

\begin{figure*}[]
\centering
\subfloat[Lennard-Jones potential, shifted by the value of the
well depth $\epsilon$. Note the single potential well, as well as the
limiting behavior of the branches $r_L$ and $r_R$ as $\phi \to \epsilon$
and $\phi \to \infty$.]
{
	\label{fig:wellwidth}
	\centering
	\includegraphics[width=0.441\textwidth]{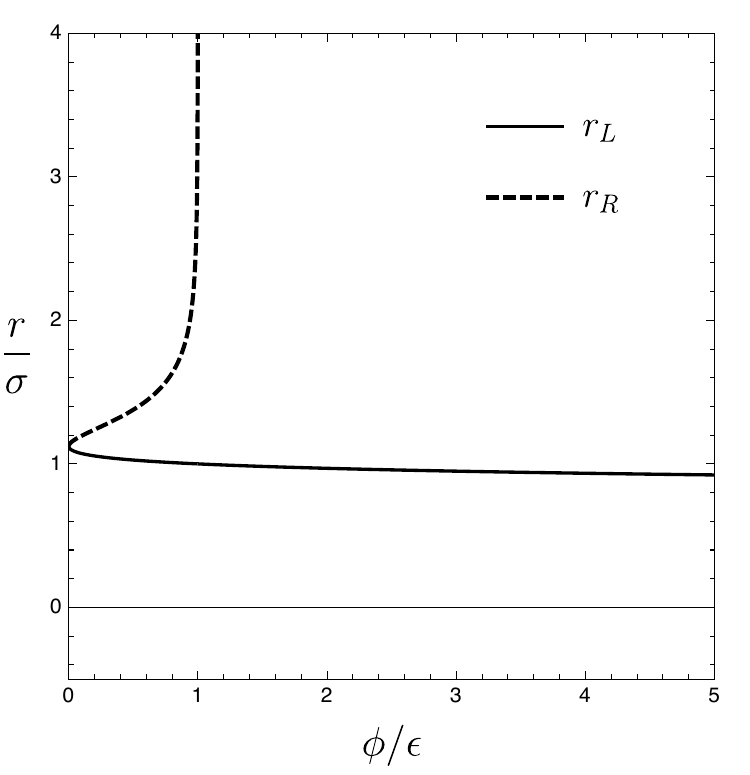}
}
\hfill
\subfloat[The well-width function $r_L^3 - r_R^3$ for the
	Lennard-Jones potential. This
	function fails to be piecewise continuous due to
	the discontinuity at $\phi = \epsilon$.]
{
	\label{fig:wellwidth}
	\centering
	\includegraphics[width=0.459\textwidth]{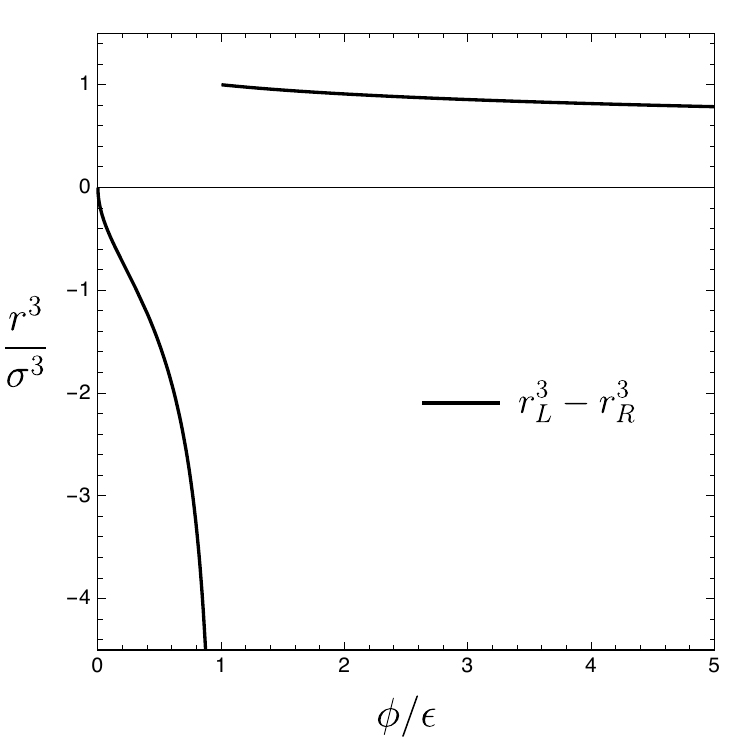}
}
\caption{The Laplace inversion procedure fails to work for
interparticle interactions~$u(r)$ with a single potential well and
the limiting behavior $u \to \infty$ as
$r \to 0$ and $u \to 0$ as $r \to \infty$, such as the Lennard-Jones potential.
This is because these assumptions imply a well-width function $r_L^3 - r_R^3$
that is not piecewise continuous, and so the Laplace transform $L(r_L^3 - r_R^3)$
is undefined.}
\end{figure*}

We begin by reviewing the inversion procedure.
Suppose that $u(r)$ possesses a single potential well of depth~$\epsilon$,
and define a shifted potential $\phi(r) = u(r) + \epsilon$, so that $\phi(r) > 0$ 
everywhere. Then $\phi(r)$ can be divided into
two monotonic branches to the left and right of the minimum; denote the inverses
of $\phi(r)$ on these by $r_L(\phi)$ and $r_R(\phi)$, respectively. 
By changing integration variable to $\phi$ on these intervals and defining
$r_R = 0$ for $\phi > \epsilon$, we can use Eq.~\eqref{b2-uderiv} to obtain
\begin{equation}
B_2(\beta) = \frac{2\pi\beta}{3}\,e^{\beta\epsilon}\int_0^\infty (r_L^3-r_R^3)
e^{-\beta \phi}\,d\phi.
\label{eq:app-B2well}
\end{equation}
At this point, it is claimed that the integral on the right
defines the Laplace transform of the ``well-width'' function
$r_L^3 - r_R^3$, so that in particular,
\begin{equation}
	r_L^3 - r_R^3 = \frac{3}{2\pi}L^{-1}
	\paren{\frac{B_2(\beta)e^{-\beta\epsilon}}{\beta}}.
	\label{eq:general-jonahrowlinson}
\end{equation}
This is stronger than the result presented in Eq.~\eqref{eq:jonahrowlinson}, since
it provides information about the difference $r_L^3 - r_R^3$ of the turning points
at a given value of the potential, and can be applied to nonmonotonic potentials.
However, there are two issues with this reasoning.
The first issue, which occurs in the derivation of Eq.~\eqref{eq:general-jonahrowlinson},
is that the well depth $\epsilon$ is neglected, either by claiming that the
zero of energy can be taken as the minimum of the potential well---2in effect,
replacing $u$  with $\phi$ in Eq.~\eqref{b2-uderiv} --
as is done in~\cite{frisch1960conditions}, or by claiming that the integral
\begin{equation}
\int_{-\epsilon}^\infty (r_L^3-r_R^3)
e^{-\beta u}\,du
\end{equation}
defines the Laplace transform of $r_L^3 - r_R^3$, as is done
in~\cite{keller1959determination}. We cannot see how either of these arguments
are valid. The expression we present in Eq.~\eqref{eq:general-jonahrowlinson}, which
was also given in~\cite{maitland1985MolPh}, correctly
accounts for this well depth. With this corrected expression, we can
use the shifting theorem~\cite{schiff2013laplace}
\begin{equation}
	L^{-1}\paren{e^{-as}G(s)}
	       = H(t-a)g(t-a),\qquad a\ge 0,\label{eq:heaviside}
\end{equation}
where $G(s) \equiv L\paren{g(t)}$ and  $H$ is the Heaviside function,
which in Eq.~\eqref{eq:general-jonahrowlinson} gives
\begin{align}
r_L^3 - r_R^3
&= \frac{3}{2\pi}H(\phi - \epsilon)\,L^{-1}\paren{\frac{B_2(\beta)}{\beta}}
(\phi - \epsilon) \nonumber\\[5pt]
&\equiv \frac{3}{2\pi}H(u)\,L^{-1}\paren{\frac{B_2(\beta)}{\beta}}(u).
\end{align}
Thus, this inversion procedure asserts that $r_L^3 - r_R^3 = 0$ for $u < 0$, a
contradiction. Note that there is no contradiction when the potential is
monotonic, for then $u > 0$ everywhere, and there is only one branch
of the potential.

The second issue concerns whether $r_L^3 - r_R^3$ has a well-defined
Laplace transform. To see this, we consider the continuity of the well-width
function. In Fig.~\ref{fig:wellwidth}, we exhibit this function
for the Lennard-Jones potential, defined by
\begin{equation}
u(r) = 4\epsilon \bkt{\paren{\frac\sigma r}^{12} - \paren{\frac\sigma r}^6},
\end{equation}
which has a well depth of $\epsilon$.
Note the discontinuity at $\phi = \epsilon$, which is typical for all
functions with a single potential well and the limiting behavior described
in the paper. This presents a difficulty for the Laplace inversion method,
because the Laplace transform is only defined for functions
that are piecewise continuous~\cite{schiff2013laplace},
which fails to be true for the well-width function.
However, as in the previous discussion, there is no difficulty for monotonic potentials.
Thus, when taken together, these results suggest that the Laplace inversion procedure can
only be applied to monotonic potentials. This justifies our assumption of
monotonicity in the derivation of Eq.~\eqref{eq:monotonicB2}.

Next, we consider the claim by Maitland \emph{et al.}~\cite{maitland1985MolPh}
that it is possible to uniquely determine the intermolecular potential
from the second virial coefficient, even for nonmonotonic potentials.
Their argument is as follows:
by elementary calculus, one can show from Eq.~\eqref{eq:app-B2well} that
\begin{equation}
	B_2 - \beta \dfr{B_2}\beta
	= -\frac{2\pi\beta^2}{3}\, e^{\beta\epsilon} \int_0^\infty (\phi - \epsilon)
	(r_L^3 - r_R^3)\, e^{-\beta\phi}
	\,d\phi.
	\label{eq:maitland}
\end{equation}
Recognizing that the right-hand side defines a Laplace transform, we obtain
\begin{equation}
\phi - \epsilon = 
-\frac1{r_L^3 - r_R^3}\frac{3}{2\pi}L^{-1}
\bkt*[\bigg]{\,\frac1{\beta^2}
\paren{B_2 - \beta \dfr {B_2}\beta}e^{-\beta\epsilon}\,}.
\label{eq:maitland-laplace}
	\end{equation}
The claim in~\cite{maitland1985MolPh} is that since $r_L^3 - r_R^3$ is given by
Eq.~\eqref{eq:general-jonahrowlinson},
it follows that Eq.~\eqref{eq:maitland-laplace}
specifies the entire potential $u \equiv \phi - \epsilon$ for a given second
virial coefficient $B_2$. The authors therefore argue that complete inversion is
possible \emph{in principle,} and that the only practical barrier is the
experimental difficulty associated with making sufficiently precise measurements.

However, due to the properties of Laplace transforms,
Eq.~\eqref{eq:maitland-laplace} is always vacuously satisfied, and therefore
cannot be used to gain new information about the potential.
To see this, observe that
\begin{align}
	&-\frac1{r_L^3 - r_R^3}\frac3{2\pi}L^{-1}
\bkt*[\bigg]{\,\frac1{\beta^2}
\paren{B_2 - \beta \dfr {B_2}\beta}e^{-\beta\epsilon}\,}\nonumber\\[5pt]
	&\hspace{80pt}{}= -\frac{\displaystyle L^{-1}
\vphantom{\Bigg(}\bkt*[\bigg]{\,\frac1{\beta^2}\paren{B_2 - \beta \dfr {B_2}\beta}
e^{-\beta\epsilon}\,}
}{\displaystyle \vphantom{\Bigg(}L^{-1}\bkt{\frac{B_2}\beta \, e^{-\beta\epsilon}}}\nonumber
\\[5pt]
&\hspace{80pt}= -\frac{-(\phi - \epsilon)f(\phi - \epsilon)}{f(\phi - \epsilon)}\nonumber \\[5pt]
&\hspace{80pt}{}= \phi - \epsilon,
\end{align}
where in the first line we substituted Eq.~\eqref{eq:general-jonahrowlinson},
and in the second line we used
 \begin{equation}
 L^{-1}\paren{\dfr[n]{}s G(s)} = (-1)^n t^n g(t), \qquad n = 1,2,3,\dots
 \label{eq:laplace-deriv}
\end{equation}
for $G(s) \equiv L(g(t))$, together with the shifting theorem
\eqref{eq:heaviside}~\cite{schiff2013laplace}.
Therefore Eq.~\eqref{eq:maitland-laplace}
is trivially satisfied, which is what we wanted to show.
The property~\eqref{eq:laplace-deriv}
shows that this is to be expected: the inverse Laplace transform of a function is
equal to the inverse transform of its derivatives up to a multiplicative factor
of $(-1)^nt^n$, so we cannot hope to obtain more information by examining
derivatives of Eq.~\eqref{eq:app-B2well}.

\bibliography{flatvirials}

\end{document}